\documentclass[twocolumn,times,tighten]{aastex701}

\begin{document}

\title{No evidence of polarization in the $11.3\,\mu$m PAH emission line by independent analyses}

\author[0000-0001-5357-6538]{Enrique Lopez-Rodriguez}
\affiliation{Department of Physics \& Astronomy, University of South Carolina, Columbia, SC 29208, USA}
\email{elopezrodriguez@sc.edu}


\begin{abstract}

Polycyclic aromatic hydrocarbons (PAHs) are commonly used as proxies for star formation, molecular gas content, and other interstellar medium (ISM) properties in our Galaxy and other galaxies. Given their abundance and brightness, polarization measurements of PAH features could, in principle, provide a probe of the ISM magnetic field and intrinsic PAH properties; however, the diagnostic power of PAH polarization remains to be established. Previous studies reported that the $11.3\,\mu$m PAH emission line in the northwestern nebula of the Herbig Be star MWC\,1080 was polarized at $1.9\pm0.2$\%. This level of polarization was explained via the paramagnetic relaxation process, which may allow the characterization of magnetic fields in the ISM. Using the same observations, here, we re-analyzed the $8-13\,\mu$m spectro-polarimetric observations taken with CanariCam on the 10.4-m Gran Telescopio CANARIAS (GTC), and we measure a polarization of $0.5\pm0.6$\% within $11.3\pm0.2\,\mu$m, consistent with an unpolarized source, $0.6\pm0.2$\% (instrumental polarization). We reproduce the previously reported polarized PAH emission line if the polarization fraction spectrum is oversubtracted by a constant instrumental polarization and the polarization uncertainties, which is inconsistent with the fundamentals of polarimetric data analysis. Thus, the published $8-13\,\mu$m spectro-polarimetric data taken with CanariCam/GTC provide no statistical evidence for a polarized $11.3\,\mu$m PAH emission line, in agreement with current dust models.

\end{abstract}



\section{Introduction} \label{SEC:INT}

Polycyclic aromatic hydrocarbons (PAHs) are carbon-based nano-sized molecules highly abundant in the interstellar medium  \citep[ISM; e.g.,][]{Draine2021}. These PAHs absorb ultraviolet (UV) and optical radiation and re-emmit it at infrared (IR) wavelengths, producing strong emission lines at $3.3$, $6.2$, $7.7$, $8.6$, $11.3$, and $12.7\,\mu$m via their stretching and bending modes \citep{Tielens2008}. Specifically for our study here, the $11.3\,\mu$m band is attributed to C-H out of plane bending modes, whose strength depends on the size, structure, charging of the PAH, and the physical conditions of the environment \citep{ATB1989,Draine2001}. Although PAH formation and evolution mechanisms are still not fully understood, PAHs account for $\sim15$\% of the interstellar carbon and $\sim20$\% of the IR power of our Galaxy and star-forming galaxies \citep{Li2020,Allamandola2021}. Thus, PAHs are typically used as proxies for star formation activity in galaxies and for the formation of molecular clouds.

The brightness of PAH emission lines makes them ideal candidates for polarimetric observations, a technique that is typically photon-starved. If these PAHs are polarized, this may open a new window for studying their intrinsic atomic properties, magnetic fields in the ISM, the $10-60$ GHz Galactic foreground in the cosmic microwave background that is thought to arise from rapidly spinning small dust grains \citep{SD2009, Draine2021}, and the question of the carrier of the 2175\,\AA~UV extinction feature. PAH polarization has been proved for since the late 80's \citep{Sellgren1988}.  Both in Davis-Greenstein \citep{DG1951} and modern radiative alignment torques (RAT) theory \citep[e.g.,][]{BG2015}, the internal alignment of the grains (aligning the spin axis with a symmetry axis of the grain) depends on paramagnetic relaxation \citep{Purcell1979}.  Paramagnetism arises when there are unpaired quantum spins in the solid, as in silicates.  Carbon solids are diamagnetic and should therefore not show efficient internal alignment, nor couple to magnetic fields.  Observationally, this is supported by the sensitive upper limits of polarization in the 3.4 $\mu$m line of aliphatic CH \citep{Chiar2006} and the centro-symmetric, radial polarization geometry of the carbon-rich envelope of the asymptotic giant branch star IRC+10216 \citep{BG2022}. Another possibility, as originally proposed by \citet{LD2000}, and elaborated on by \citet{Hoang2014}, is that very small grains, including PAHs, may be aligned via resonance paramagnetic alignment.  Recent modeling of far-UV (FUV) spectro-polarimetry (Tram et al. 2026, submitted) indicates that if active, this mechanism is sub-dominant to RAT alignment induced by sub-Lyman limit radiation in fully ionized gas.

A long-standing problem in interstellar astronomy is the carrier of the 2175\,\AA~extinction feature \citep[e.g.,][]{Whittet2003}.  Since the IR emission of PAHs are thought to require excitations by UV photons \citep[e.g.,][]{Massa2022}, and the characteristics of the 2175\,\AA~feature, in terms of elemental abundances and grain size requirements imply small, likely carbonaceous grains \citep[cf][p. 97-102]{Whittet2003}. Thus, an association between these two phenomena have been widely assumed.  Observational support for this association has recently been presented by \citet{Massa2022, Gordon2024}.  The recent detection of the 2175\,\AA~feature at high redshift \citep[z$\sim2-12$; e.g.][]{Ormerod2025, Fisher2025, Markov2025} challenges the elemental enrichment in galaxies, requiring a rapid production of carbon to account for the feature as caused by such dust.

However, in the final sample of interstellar UV polarimetry \citep{Martin1999} from the two flights of the Wisconsin Ultraviolet Photo-Polarimetry Experiment (WUPPE; \citet{Nordsieck1994, Anderson1996}) and HST/FOS \citep{Allen1993, Clayton1995}, only two lines of sight, out of 30, show detectable polarization in the 2175\,\AA~feature.  Hence, a better understanding of the polarization properties of the PAH features is of central importance also to the interpretation of the UV feature.

The $11.3\,\mu$m PAH emission was reported to be polarized, $1.9\pm0.2$\%, in the northwestern (NW) nebula of the Herbig Be star MWC\,1080  \citep{Zhang2017}. These observations used $8-13\,\mu$m spectro-polarimetric observations taken with CanariCam \citep{Telesco2003} on the 10.4-m Gran Telescopio CANARIAS (GTC). This high polarization was attributed to PAH alignment with a local magnetic field via a resonance paramagnetic relaxation process. However, this polarization measurement presents several challenges. 

From an observational point of view, a) the measured polarized $11.3\,\mu$m PAH emission is unexpectedly wider than the total-intensity emission line, which indicates some issues with the analysis of the polarized spectrum,  b) the polarization fraction spectrum exhibits several highly polarized regions and low signal-to-noise ratio across the $8-13\,\mu$m wavelength range, which may affect the polarization of the PAH line, and c) the polarization fraction spectrum shows negative values, which is incompatible with polarimetry standard procedures.

From a theoretical perspective, the measured PAH polarization level exceeds that predicted by current dust models \citep{Draine2021, Hensley2023}. Dust models predict that a polarization spectrum that peaks at wavelengths $<0.5\,\mu$m may indicate additional mass distributions of small grain sizes, $<0.02\,\mu$m \citep{KM1995}. Optical polarimetric observations of MWC\,1080 show a peak at $\sim0.7\,\mu$m in the polarization spectrum (corrected by interstellar polarization) with dust models suggesting grain sizes in the range of $0.02-0.1\,\mu$m \citep[i.e., larger than PAHs; ][]{MB2001}. In addition, a lack of UV polarization implies that small grains (nanometer-sizes) are not aligned \citep{KM1995}. Only two line-of-sight towards stars in the Milky Way have tentative evidence of polarization in the $2175$\,\AA\, \citep{Clayton1992,Wolff1993}, which has been interpreted as aligned PAHs \citep{HLM2013}. For rotational emission from spinning PAHs, the alignment mechanisms in PAHs are suppressed due to energy quantization \citep{DH2016}. Although geometrical effects \citep{SD2009} cannot be ruled out. Given the implications of detecting polarized PAH emission as a tracer of magnetic fields or radiation fields, the detection of polarization in the $11.3\,\mu$m PAH emission line by \citet{Zhang2017} must be carefully considered and independently verified.

In this manuscript, we re-reduce the $8-13\,\mu$m spectro-polarimetric observations of the NW\,MWC\,1080 nebula taken with CanariCaam/GTC, and revised the calibration polarization methods. The manuscript is organized as follows. Observations and data reduction are described in Section \ref{sec:OBS}, our results are presented in Section \ref{sec:PAH}, and we reproduce and compare results in Section \ref{sec:Z2017}. We conclude with final remarks in Section  \ref{sec:CON}.

\section{Observations and Data Reduction} \label{sec:OBS}

The spectro-polarimetric observations of the $11.3\,\mu$m PAH of the MWC\,1080 NW nebula were taken with the polarization mode of CanariCam \citep{Telesco2003} on the 10.4-m GTC. We used the same datasets as those previously analyzed by \citet{Zhang2017} taken on 2015 July 31, and August 5 and 7 (GTC3-15AFLO; PI: Telesco, C.).

CanariCam was the Mid-IR (MIR; $8-25\,\mu$m) imaging, spectroscopic, imaging- and spectro-polarimetric, and coronographic instrument on the 10.4-m GTC in La Palma, Spain. CanariCam uses a $320\times240$ pixels$^{2}$ Raytheon detector array with a pixel scale of $0\farcs079$ (Nyquist sampling at $8\,\mu$m) with a field-of-view (FOV) of $26\arcsec \times 19\arcsec$. The spectro-polarimetric mode uses a half-wave plate (HWP) retarder, a field mask, a slit, and a Wollaston prism. In standard polarimetric observations, the HWP is set to four position angles (PA) in the following sequence: $0^{\circ}$, $45^{\circ}$, $22.5^{\circ}$, and $67.5^{\circ}$. The field mask comprises three available slots, each $320\times25$ pixels$^{2}$ ($25\farcs6 \times 2\farcs0$). Spectro-polarimetric observations used the low-resolution mode (R$=\Delta\lambda/\lambda=50$) with a slit width of $1\farcs04 \times 2\farcs08$ and the long axis oriented at $45^{\circ}$ East of North (E of N) at the position RA(J2000)= +23:17:26.390 and DEC(J2000) = +60:50:51.34. Observations were made using the chop-nod observing technique to remove time-variable sky background and telescope thermal emission, and to reduce the effect of $1/f$ noise from the array and electronics. The chop-throw was $11\arcsec$, the chop-angle was $-45^{\circ}$ E of N, and the chop-frequency was 1.38 Hz. The angle of the short axis of the array with respect to north on the sky (i.e., instrumental position angle, IPA) was $45^{\circ}$ E of N, and the telescope was nodded every $46.3$ s along the chopping direction. Only one slot of the field mask was used for observations, leaving the other two for noise estimation. The unpolarized star HD\,213310 and the polarized young stellar object (YSO) AFGL\,2591 \citep{Smith2000} were observed with the same instrumentation configuration immediately before or after the science observations.

Data were reduced using custom \textsc{python} routines following the same prescriptions successfully used in previous spectro-polarimetric observations of active galactic nuclei with CanariCam \citep{LR2016,LR2018}.  We describe the data reduction steps in detail. 

We estimate the difference for each chopped pair and nod frame, and then difference them for each HWP PA. All nods were examined for high or variable background that could indicate the presence of variable precipitable water vapour. No data was removed for any of these reasons. Then, we sum all chop-nod-corrected files for each HWP PA. Next, the ordinary (o-ray) and extraordinary (e-ray) rays, produced by the Wollaston prism, were identified as the bottom and top spectra for each HWP PA. Note that the identification of the o-ray and e-ray is arbitrary; the only requirement is to be consistent with the label of each of them for each HWP PA. The center of the mask was identified as the peak of the spectra in the mask. Each o- and e-ray spectrum is extracted using a width of $21$ pixels ($1\farcs60$) as performed by \citet{Zhang2017}.

The Stokes parameters, I, $q$ ($q=Q/I$), and $u$ ($u=U/I$), are estimated using the double method ratio described by \citet{Tinbergen2005}, such as

\begin{equation}
q = \frac{R_{q} -1}{R_{q} + 1};~~ \\
u = \frac{R_{u} -1}{R_{u} + 1}
\end{equation}
\noindent
where, 

\begin{equation}
R_{q}^{2} = \frac{o_{0}/e_{0}}{o_{45}/e_{45}};~~ \\
R_{u}^{2} = \frac{o_{22.5}/e_{22.5}}{o_{67.5}/e_{67.5}} \\
\end{equation}
\noindent
where $o_{PA}$ and $e_{PA}$ are the o-ray and e-ray spectra for the HWP PAs ($PA=0^{\circ}, 45^{\circ}, 22.5^{\circ}, 67.5^{\circ}$).

The total intensity, I, is estimated as

\begin{equation}\label{eq:I}
I = \frac{1}{4}[\sum_{PA}(I_{o_{PA}} + I_{e_{PA}})]
\end{equation}

The polarization fraction, $P_{\rm{b}}$, and angle $\theta$, are estimated as

\begin{equation}\label{eq:Pfrac}
P_{\rm{b}} = \sqrt{q^{2} + u^{2}};~~\theta=\frac{1}{2}\arctan(\frac{u}{q})
\end{equation}

The uncertainties of the polarization fraction, $\sigma_{p}$, and angle, $\sigma_{\theta}$, are estimated using propagation of errors, such as

\begin{equation}
\sigma_{p} = \sqrt{\frac{(q \sigma_{q})^2 + (u \sigma_{u})^2}{q^2 + u^2}}
\end{equation}

\begin{equation}
\sigma_{\theta} = \frac{1}{2} \sqrt{\frac{(u \sigma_{q})^2 + (q  \sigma_{u})^2}{q^2 + u^2} }
\end{equation}
\noindent
where the uncertainties of the Stokes parameters $\sigma_{q}$ and $\sigma_{u}$ are estimated as

\begin{widetext}
\begin{eqnarray}
\sigma_{q} = \frac{R_{q}}{(R_{q} +1)^2} \sqrt{ (\frac{\sigma_{o_{0}}}{o_{0}})^2 + (\frac{ \sigma_{e_{45}}}{e_{45}})^2 + (\frac{\sigma_{e_{0}}}{e_{0}})^2 + (\frac{\sigma_{0_{45}}}{e_{45}^2})^2 } \\
\sigma_{u} = \frac{R_{u}}{(R_{u} +1)^2} \sqrt{ (\frac{\sigma_{o_{22.5}}}{o_{22.5}})^2 + (\frac{ \sigma_{e_{67.5}}}{e_{67.5}})^2 + (\frac{\sigma_{e_{22.5}}}{e_{22.5}})^2 + (\frac{\sigma_{0_{67.5}}}{e_{67.5}^2})^2 } 
\end{eqnarray}
\end{widetext}
\noindent
where $\sigma_{o_{PA}}$ and $\sigma_{e_{PA}}$ are the uncertainties of the o-ray and e-ray spectra for each HWP PA. These uncertainties were estimated as the standard deviations within the aperture width for each pixel across the wavelength axis of the empty slots in the mask.

\begin{figure}[ht!]
\includegraphics[width=\columnwidth]{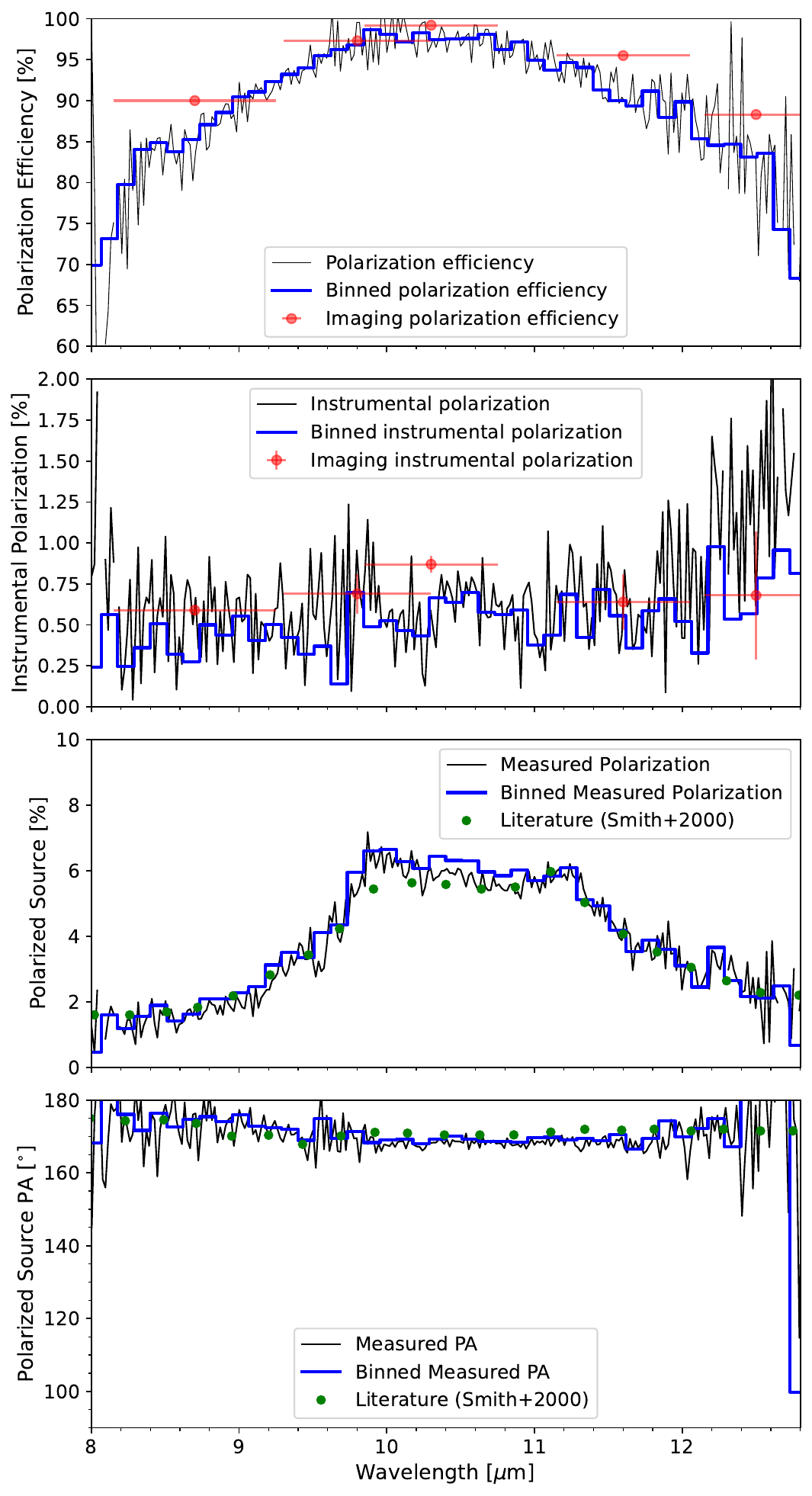}
\caption{Calibration. The polarization efficiency (top), instrumental polarization (second row), and polarization (third row) and PA (bottom) of the polarized source, AFGL\,2591, at the original resolution (black), $6$ spectral pixels ($0.12$ $\mu$m) binned (blue) are shown. For comparison, the polarization measurements using CanariCam narrow-band imaging filters (red circles) and the spectro-polarimetric observations (green circles) from \citet{Smith2000} are displayed.
\label{fig:fig1}}
\end{figure}

The polarization fraction, $P_{\rm{b}}$ (Eq. \ref{eq:Pfrac}), is a positive quantity that needs to be debiased \citep{WK1974}, as

\begin{equation}\label{eq:Pdebias}
P = \sqrt{P_{\rm{b}}^2 -\sigma_{p}^2} = \sqrt{q^2 + u^2 - \sigma_{p}^2}
\end{equation}

We correct the measured Stokes parameters for the polarization efficiency and the instrumental polarization. The polarization efficiency spectrum was taken from the CanariCam archive and previously used in the data reduction of active galaxies by \citet{LR2016,LR2018}. The polarization efficiency spectrum was obtained by observing the unpolarized star HD\,186791 on 2013 July 21 through a wire-grid before the entrance window of CanariCam. The polarization efficiency is shown in Figure \ref{fig:fig1}. The instrumental polarization spectrum was estimated using the unpolarized star, HD\,213310, observed before or after the science target. Figure \ref{fig:fig1} shows the instrumental polarization. Both the polarization efficiency and instrumental polarization are consistent with the quoted values by the GTC\footnote{CanariCam details at GTC \url{https://www.gtc.iac.es/instruments/canaricam/canaricam.php}}. For comparison, we show the measurements of the instrumental polarization and polarization efficiency in the imaging filters: Si2-8.7 ($\lambda_{c} = 8.7\,\mu$m, $\Delta\lambda = 1.1\,\mu$m, $50$\% cut-off), Si3-9.8 ($\lambda_{c} = 9.8\,\mu$m, $\Delta\lambda = 1.0\,\mu$m), Si4-10.3 ($\lambda_{c} = 10.3\,\mu$m, $\Delta\lambda = 0.9\,\mu$m), Si5-11.6 ($\lambda_{c} = 11.6\,\mu$m, $\Delta\lambda = 0.9\,\mu$m), and Si6-12.5 ($\lambda_{c} = 12.5\,\mu$m, $\Delta\lambda = 0.7\,\mu$m).

The final Stokes parameters, corrected by polarization efficiency and instrumental polarization, of the polarized YSO and science target are estimated as

\begin{equation}\label{eq:qu_Pinst_Peff}
q = \frac{q-q_{\rm{inst}}}{q_{\rm{eff}}} ;~~u=\frac{u-u_{\rm{inst}}}{u_{\rm{eff}}}
\end{equation}
\noindent
where $q_{\rm{inst}}$ and $u_{\rm{inst}}$ are the Stokes parameters of the instrumental polarization spectrum, and $q_{\rm{eff}}$ and $u_{\rm{eff}}$ are the Stokes parameters of the polarization efficiency spectrum.

The polarized YSO, AFGL\,2591, was observed to calibrate the polarization angle of the science source, MWC\,1080\,NW nebula. In addition, we use AFGL\,2591 to show the quality of our data reduction steps. We show the polarized spectrum of AFGL\,2591 in Figure \ref{fig:fig1}, which is in excellent agreement with the polarization fraction and angle spectra measured by \citet{Smith2000}. The polarization fraction and angle spectra of AFGL\,2591 were digitalized from \citet{Smith2000} using WebPlotDigitilizer\footnote{WebPlotDigitizer can be found at \url{https://automeris.io/}}. The polarization angle calibration offset between the observations and the measurements by \citet{Smith2000} is $170^{\circ}$. However, this angular offset is not included in any of the data analyses that affect the polarization of the $11.3\,\mu$m PAH line.

\begin{figure*}[ht!]
\includegraphics[width=\textwidth]{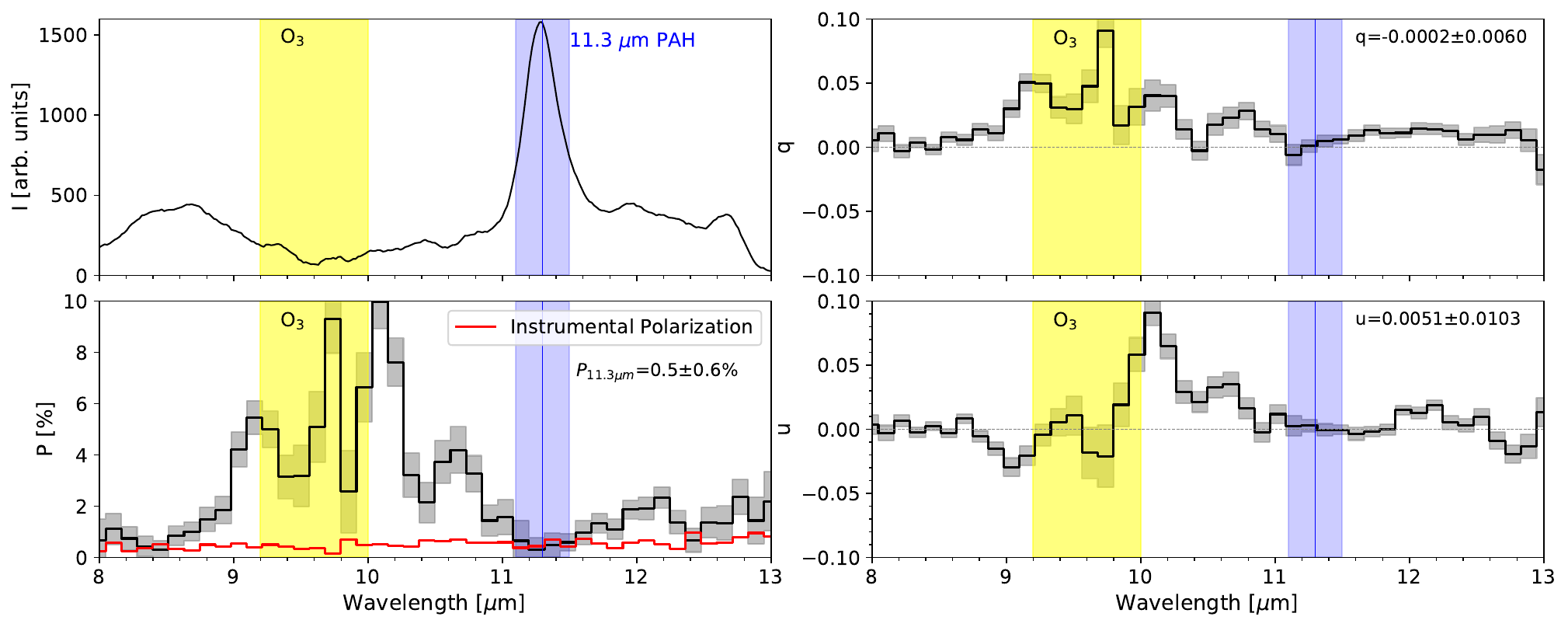}
\caption{Spectro-polarimetric observations of MWC\,1080\,NW. The Stokes I (top-left), $q$ (top-right), and $u$ (bottom-right), the polarization fraction (bottom-left) spectra (black line), and their $1\sigma$ uncertainties (gray region) are shown. The instrumental polarization (red line; as in Fig. \ref{fig:fig1}) is displayed in the same panel as the polarization fraction spectrum. The Stokes I has been smoothed using a $0.06\,\mu$m boxcar, while the Stokes $q$ and $u$ are downsampled to $0.12\,\mu$m and then smoothed with a $0.36\,\mu$m boxcar. The $11.3\,\mu$m PAH emission (blue solid line), the aperture width of $0.4\,\mu$m (blue region), and the approximated extent of the telluric $0_{3}$ band (yellow region) are shown. The values of the Stokes $q$ and $u$ and the polarization fraction and their uncertainties, within the $11.3\pm0.2\,\mu$m wavelength range (blue region), are displayed.
\label{fig:fig2}}
\end{figure*}

As performed by \citet{Zhang2017}, to increase the signal-to-noise ratio (SNR) of the spectra, we bin the o-ray and e-ray spectra by summing the counts within bins of $6$ pixels ($0.12\,\mu$m), and further applied a $3$ pixels ($0.36\,\mu$m) box-smoothing. The binned o-ray and e-ray spectra are then used following the procedure described above. For completeness, we show the binned spectra of the instrumental polarization, polarization efficiency, and polarized YSO in Fig. \ref{fig:fig1}. In all cases, the binned spectra are consistent with the original spectra.

\section{The $11.3\,\mu$m PAH emission line is unpolarized} \label{sec:PAH}

\subsection{Total spectra}\label{subsec:TS}

Figure \ref{fig:fig2} shows the $8-13\,\mu$m spectro-polarimetric observations of the MWC\,1080\,NW nebula. The total intensity spectrum (top-left) shows the detection of the $11.3\,\mu$m PAH emission line. The Stokes parameters $q$ (top-right) and $u$ (bottom-right) and the polarization fraction (bottom-left) with the $1\sigma$ uncertainties (gray shadow) are shown. We show the approximated extent of the telluric $0_{3}$ band (yellow region) for all spectra. As performed by \citet{Zhang2017}, the spectra are extrated within a $1\farcs6$ ($21$ pixels), and the Stokes I is smoothed using a $3$ pixels ($0.06\,\mu$m) boxcar and the Stokes $q$ and $u$ are downsampled to 6 pixels ($0.12\,\mu$m) and, then, smoothed with a $3$ pixels ($0.36\,\mu$m) boxcar. 

Figure \ref{fig:fig2} shows the central location (blue line) and the full-width-at-half-maximum (FWHM) of $\sim0.4\,\mu$m (blue region). This region contains the lowest polarization signal within the polarization spectra. Specifically, we measure a polarization fraction of $0.5\pm0.6$\% within $11.3\pm0.2\,\mu$m. We show the instrumental polarization spectra (red line), which has the same polarization fraction level, $0.5\pm0.2$\% as the $11.3\,\mu$m PAH emission line. The Stokes $q$ ($1\times10^{-4}$) and $u$ ($5\times10^{-3}$) also show values consistent with an unpolarized source. Thus, we measure the $11.3\,\mu$m PAH emission line to be consistent with an unpolarized source.

\subsection{Continuum subtracted spectra}\label{subsec:CS}

\begin{figure*}[ht!]
\includegraphics[width=\textwidth]{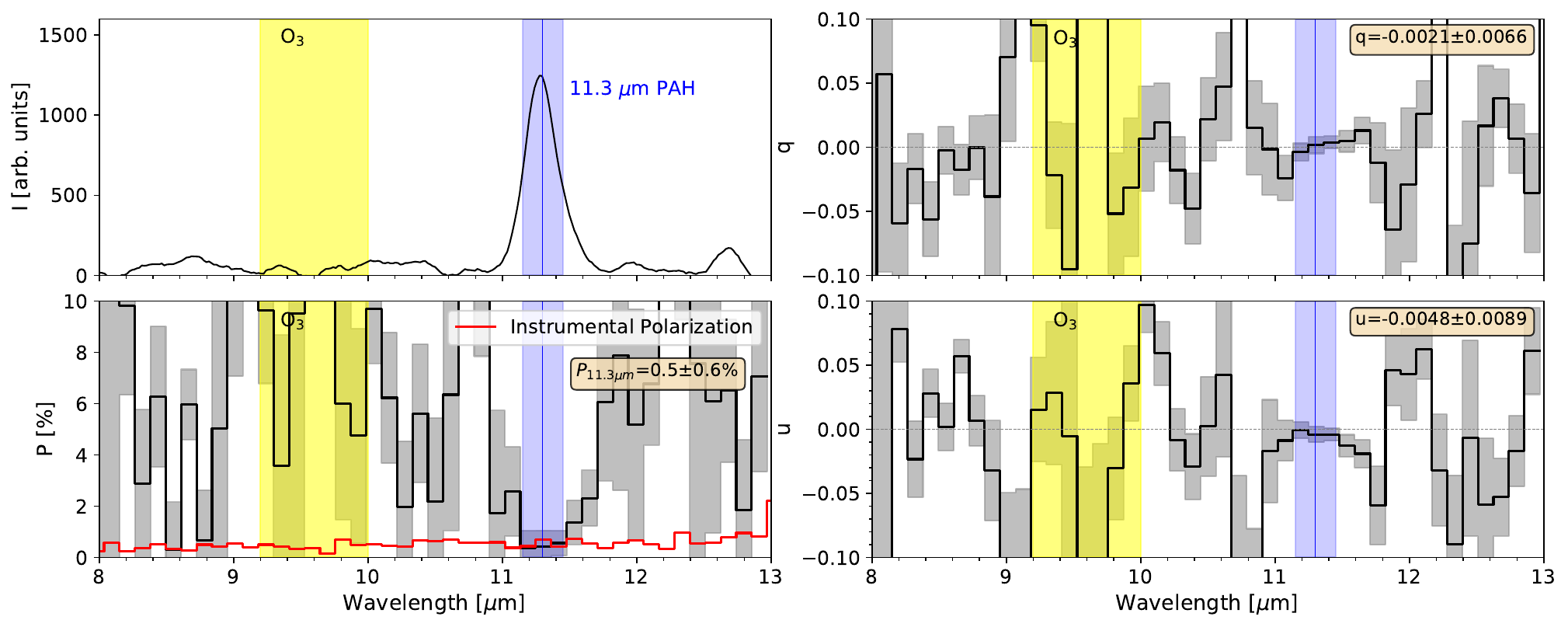}
\caption{Continuum subtracted spectro-polarimetic observations of MWC\,1080\,NW. Same display as Fig. \ref{fig:fig2} but for the continuum-subtracted spectro-polarimetric observations. 
\label{fig:fig3}}
\end{figure*}

The continuum and the $11.3\,\mu$m PAH emission line may have different polarization fractions and angles. This difference in polarization properties may cancel out the intrinsic polarization of the PAH line. To investigate this effect, we subtract the continuum component in the o-ray and e-ray spectra for each HWP PA. 

To subtract the continuum emission, the o-rays and e-rays spectra for each HWP PA are fit with a polynomial function of order eight within the $8-13\,\mu$m wavelength range, where the $11.3\pm0.5\,\mu$m wavelength range is masked. The fitting and continuum subtracted spectra are shown in Appendix \ref{app:app1} (Figure \ref{fig:App1_fig1}). The final Stokes I, $q$, and $u$, and the polarization fraction spectra are estimated using the same methodology as above. Figure \ref{fig:fig3} shows that the $11.3\,\mu$m PAH emission line has a polarization fraction of $0.5\pm0.6$\% within $11.3\pm0.2\,\mu$m, which is consistent with an unpolarized source. Outside of this wavelength range, the polarization spectra are dominated by noise, as expected by the continuum background-subtracted spectra.

\section{Reproduction of Zhang+2017 results}\label{sec:Z2017}

\begin{figure*}[ht!]
\includegraphics[width=\textwidth]{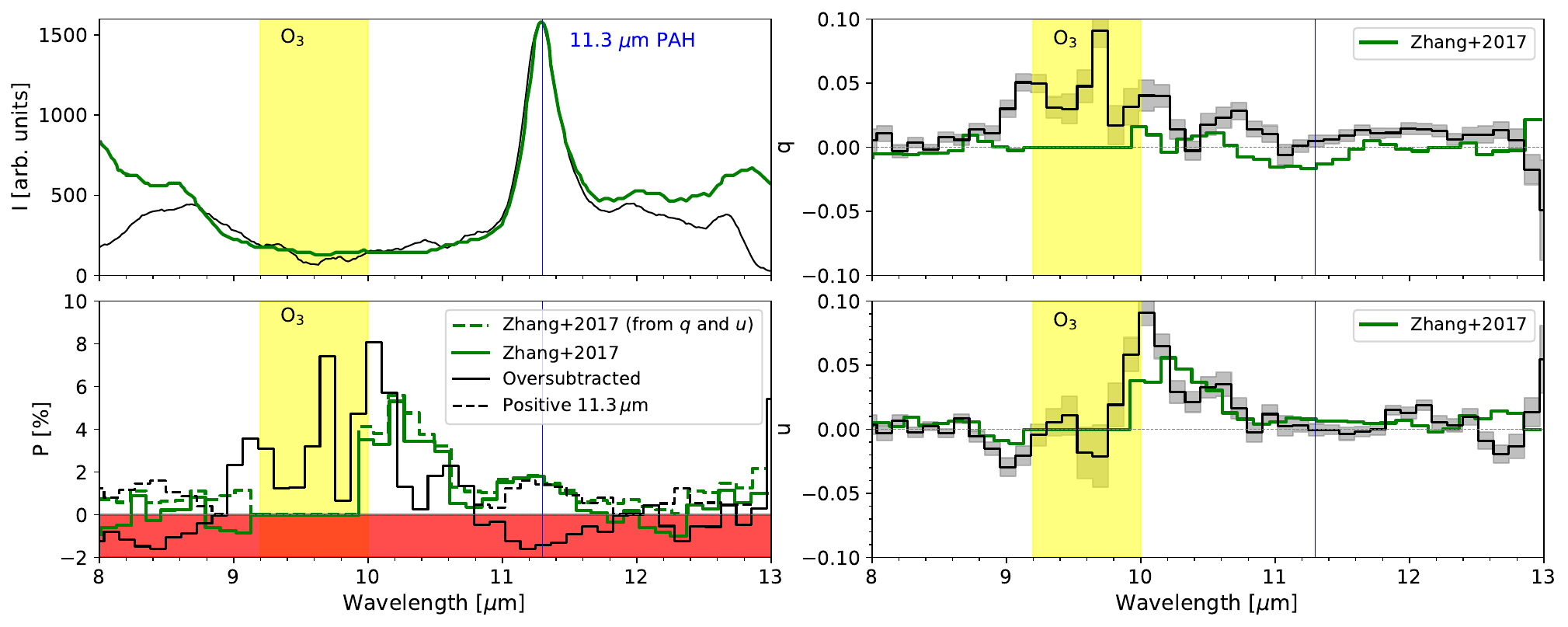}
\caption{Comparison with and reproduction of \citet{Zhang2017} results. 
Our Stokes I, $q$, and $u$ (black lines) are the same as in Fig. \ref{fig:fig1}. 
Our polarized spectrum, without debias ($P_{\rm{b}}$), were subtracted by a constant instrumental polarization of $0.6$\%, and the uncertainty of the polarization fraction, $\sigma_{p}$, such as $P=P_{\rm{b}} - P_{\rm{inst}} - \sigma_{\rm{p}}$ (black solid line). The absolute values of this polarization spectra are display as solid dashed lines. 
The digitalized spectra of \citet{Zhang2017} are shown as green solid lines. 
The polarization fraction spectra estimated using the Stokes $q$ and $u$ is displayed as dashed green lines.
\label{fig:fig4}}
\end{figure*}

To understand the origin of the measured polarized $11.3\,\mu$m PAH emission reported by \citet{Zhang2017}, we reproduce and compare our results with theirs. We digitalized their figures 2 and 3 using WebPlotDigitizer\footnote{WebPlotDigitizer can be found at \url{https://automeris.io/}}, which extracts the Stokes I, $q$, and $u$, and the polarization fraction spectra within the $8-13\,\mu$m wavelength range. Their spectra (green lines) are shown in Figure \ref{fig:fig4}. We find several similarities and differences.

The $11.3\,\mu$m PAH emission line is reproducible, however the Stokes I shows substantial differences within the $8-8.8\,\mu$m and $11.8-13\,\mu$m wavelength ranges. Our total fluxes decrease at shorter and longer wavelengths, whereas \citet{Zhang2017} reports an increase in total flux. The o-ray and e-rays for each HWP PA extracted within the same aperture as \citet{Zhang2017} are shown in Fig. \ref{fig:App1_fig1} (Appendix \ref{app:app1}), which are used to estimate the Stokes I (Eq. \ref{eq:I}). For all o-ray and e-ray spectra, the fluxes show a decrease in flux at the edges of the spectra (i.e., $8-8.8\,\mu$m and $11.8-13\,\mu$m). Individual observations of MWC\,1080\,NW were inspected, but the same behavior was found. We are not able to reproduce the increase of fluxes at the edges of the total intensity spectrum. However, these wavelength ranges do not affect the characterization of the  $11.3\,\mu$m PAH emission line, whose total intensity is reproduced by our data reduction.

The major inconsistencies with the polarization definitions are found in their polarization fraction spectrum. Their polarization fraction spectrum shows negative values within the $8-8.8\,\mu$m and $11.8-13\,\mu$m wavelength ranges (Fig. \ref{fig:fig4}). This result cannot be reproduced because, mathematically, the polarization fraction is a positive quantity, $P_{\rm{b}} = \sqrt{q^2+u^2}$ (Eq. \ref{eq:Pfrac}). Furthermore, the polarization fraction is debiased by computing $P = \sqrt{q^2+u^2-\sigma_{p}^2}$ (Eq. \ref{eq:Pdebias})---also a positive quantity. However, \citet{Zhang2017} quoted: \textit{"Debiasing may introduce negative values if the noise fluctuations are stronger than the signal"} in their section 2, which may indicate a potential overestimation of the polarization fraction in their data analysis. Negative values inside the square root indicate that the polarization fraction is dominated by noise and, therefore, this polarization measurement is not statistically significant. We investigate this further.

We reproduce the polarized $11.3\,\mu$m PAH emission line by over-subtracting the polarization fraction without using the square root in Eq. \ref{eq:Pfrac} (Fig. \ref{fig:fig4}). Specifically, the oversubtracted polarization fraction is computed as $P= P_{\rm{b}}-P_{\rm{inst}}-\sigma_{p}$---this may be the only mathematical expression to obtain negative polarization fractions---, where $P_{\rm{inst}} = 0.6$\% is the instrumental polarization and $\sigma_{p}$ is the uncertainty of the polarization fraction. We subtract a constant $P_{\rm{inst}}$ across the $8-13\,\mu$m wavelength range as performed by \citet{Zhang2017}, rather than the subtraction of the instrumental polarization spectrum shown in Fig. \ref{fig:fig1} using Eqs. \ref{eq:qu_Pinst_Peff}. Note that it is unclear whether polarization efficiency was applied by \citet{Zhang2017}; we apply it here. Then, the absolute value of the oversubtracted polarization fraction is computed, which produces an artificial polarized line at $11.3\,\mu$m. Figure \ref{fig:fig4} shows the reproduced polarization fraction spectra (black dashed line) with the $11.3\,\mu$m PAH emission polarization estimated to be $1.4\pm0.6$\%, in agreement with that reported, $1.9\pm0.2$\%, by \citet{Zhang2017}. The slight differences in polarization fraction may be due to the error propagation and the correction of the polarization efficiency; however, both results are compatible within their uncertainties.

We further note that using the Stokes $q$ and $u$ spectra from \citet[][fig. 4]{Zhang2017}, the resulting polarization fraction is positive across the full wavelength range (green dashed line). Thus, their reported polarization fraction spectrum \citep[green solid line; fig. 2 by][]{Zhang2017} and the one computed from their Stokes $q$ and $u$ (green dashed line) are inconsistent. For completeness, we show them both in Figure \ref{fig:fig4}. Specifically, the polarized spectrum is negative in the $8-9\,\mu$m and $11.8-12.5\,\mu$m wavelength range. As previously noted, the polarized spectrum must be a positive quantity. We hypothesize that the reported Stokes $q$ and $u$ from \citet{Zhang2017} may need to be estimated from the oversubtracted $P$ spectrum and the PA spectrum. This is computed by solving the set of equations shown in Eqs. \ref{eq:Pfrac}. This computation will be consistent with the artificial $11.3\,\mu$m PAH line. Note that our `positive polarization spectrum' (black dashed line) is compatible with the polarized spectrum computed from their Stokes $q$ and $u$ (green dashed line), showing that our suggested oversubtracted polarization spectrum is the main reason for the artificial $11.3\,\mu$m PAH line and that the negative values are artificially produced by the oversubtracted polarization procedure shown above.

\section{Final Remarks} \label{sec:CON}

We re-analyzed the $8-13\,\mu$m spectro-polarimetric observations of the MWC\,1080\,NW nebula taken with CanariCam/GTC to investigate the reported polarized emission of the $11.3\,\mu$m PAH emission line by \citet{Zhang2017}. Our independent analysis shows that the $11.3\,\mu$m PAH emission has a polarization fraction of $0.5\pm0.6$\%, which is consistent with the instrumental polarization, $0.6\pm0.2$\%. This result is obtained in both the total emission spectra and the continuum-subtracted spectra. We reproduce \citet{Zhang2017} results if the polarization fraction spectrum is over-subtracted by a constant instrumental polarization and the polarization uncertainties.  However, this oversubtraction is inconsistent with the standard definitions of polarization analysis. The report of unpolarized $11.3\,\mu$m PAH is consistent with dust models \citep[e.g.,][]{SD2009,Draine2021,Hensley2023} and previously reported unpolarized $11.3\,\mu$m PAH emission in the active galaxies, NGC\,1068 and NGC\,4151 \citep{LR2016,LR2018}. Deeper spectro-polarimetric observations in star-forming regions \citep[e.g.,][]{Peeters2004} in our galaxy and in nearby galaxies \citep[e.g.,][]{Egorov2025} and starbursts \citep[e.g.,][]{Villanueva2025,Lopez2025}  will be of great interest to establish the polarization expectations of PAH lines.

\begin{acknowledgments}
E.L.R. thanks the anonymous referee for the positive and useful feedback. 
E.L.R. thanks B.G. Andersson and Brandon Hensley for their insights into the polarization mechanisms of PAHs. This work is based on data from the GTC Public Archive at CAB (INTA-CSIC), developed in the framework of the Spanish Virtual Observatory project supported by the Spanish MINECO through grants AYA 2011-24052 and AYA 2014-55216. The system is maintained by the Data Archive Unit of the CAB (INTA-CSIC).

\end{acknowledgments}

\begin{contribution}

E.L.R. performed data reduction, analysis, interpretation of results, writing, and submission of the manuscript. 

\end{contribution}

%
\facilities{GTC(CanariCam)}

\software{astropy \citep{astropy:2013,astropy:2018,astropy:2022}
          }


\appendix

\section{Continuum subtraction}\label{app:app1}

\begin{figure*}[ht!]
\includegraphics[width=\textwidth]{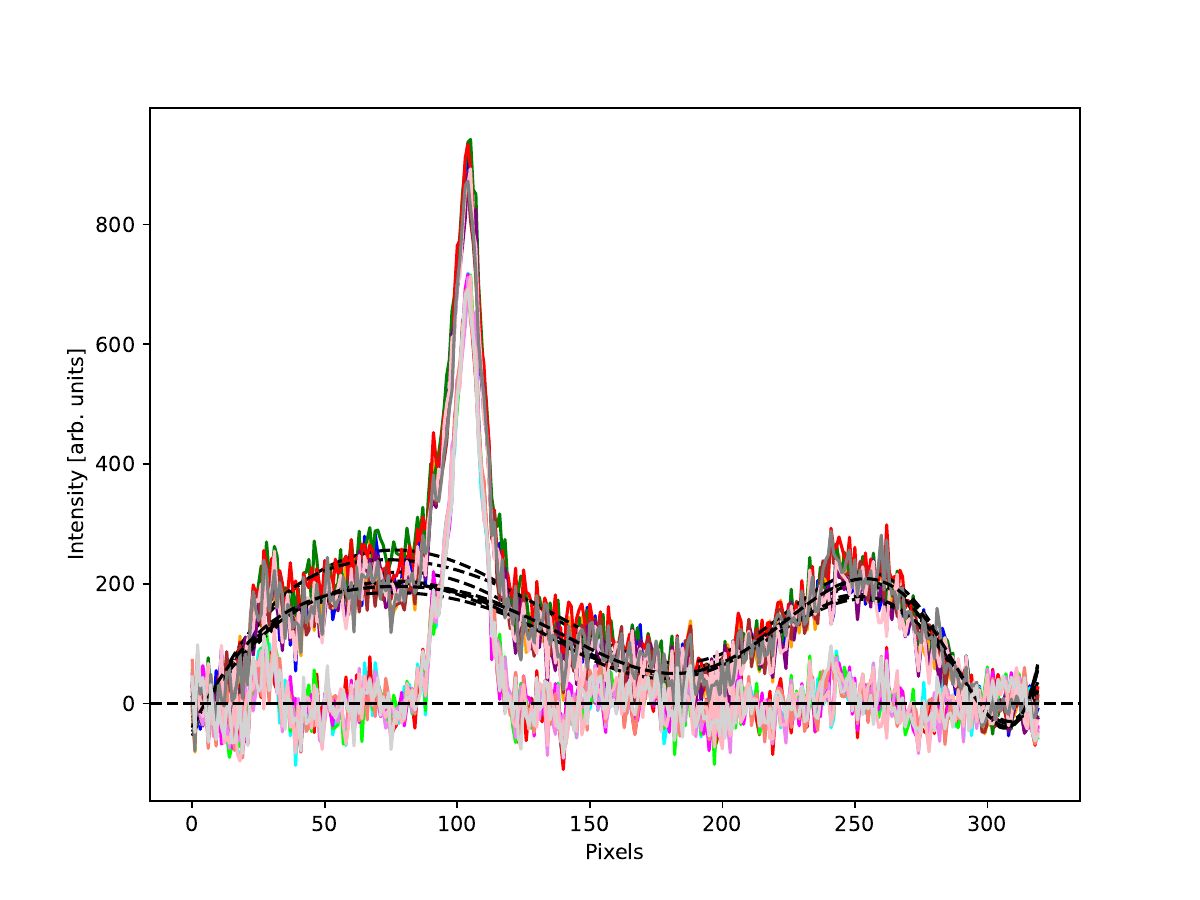}
\caption{Continuum subtraction of the o-rays and e-rays spectra. The o-rays and e-rays spectra, the continuum fitting lines (dashed line), and the continuum-subtracted spectra are shown. 
\label{fig:App1_fig1}}
\end{figure*}

Figure \ref{fig:App1_fig1} shows the extracted o-rays and e-rays for each HWP PA, the continuum fit (dashed line), and the continuum-subtracted spectra. 

\bibliography{references}{}
\bibliographystyle{aasjournalv7}



\end{document}